\documentclass[preprint,prd,nofootinbib,onecolumn,showkeys,showpacs]{revtex4}%
\usepackage{amsfonts}
\usepackage{amsmath}
\usepackage{amssymb}
\usepackage{graphicx}
\usepackage{rotating}
\usepackage{float}%
\providecommand{\U}[1]{\protect\rule{.1in}{.1in}}

\ifx\pdfoutput\relax\let\pdfoutput=\undefined\fi
\newcount\msipdfoutput
\ifx\pdfoutput\undefined\else
\ifcase\pdfoutput\else
\ifx\paperwidth\undefined\else
\ifdim\paperheight=0pt\relax\else\pdfpageheight\paperheight\fi
\ifdim\paperwidth=0pt\relax\else\pdfpagewidth\paperwidth\fi
\fi\fi\fi
\begin{document}
\preprint{ }
\title[Conformal Cosmology - Part III]{Conformal Cosmology and the Pioneer Anomaly}
\author{Gabriele U. Varieschi}
\affiliation{Department of Physics, Loyola Marymount University - Los Angeles, CA 90045,
USA\footnote{E-mail: Gabriele.Varieschi@lmu.edu}}

\begin{abstract}
We review the fundamental results of a new cosmological model, based on
conformal gravity, and apply them to the analysis of the early data of the
Pioneer anomaly.

We show that our conformal cosmology can naturally explain the anomalous
acceleration of the Pioneer 10 and 11 spacecraft, in terms of a local
blueshift region extending around the solar system and therefore affecting the
frequencies of the navigational radio signals exchanged between Earth and the
spacecraft. On the contrary, conformal gravity corrections alone would not be
able to account for dynamical effects of such magnitude to be capable of
producing the observed Pioneer acceleration.

By using our model, we explain the numerical coincidence between the value of
the anomalous acceleration and the Hubble constant at the present epoch and
also confirm our previous determination of the cosmological parameters
$\gamma\sim10^{-28}\ cm^{-1}$ and $\delta\sim10^{-4}-10^{-5}$. New Pioneer
data are expected to be publicly available in the near future, which might
enable more precise evaluations of these parameters.

\end{abstract}
\startpage{1}
\endpage{ }
\maketitle
\tableofcontents

\section{\label{sect:introduction}Introduction}

The Pioneer 10 and 11 spacecraft were launched in the early 1970s, to conduct
explorations in the region of the solar system beyond the orbit of Mars and to
perform close observations of Jupiter. They were also the first spacecraft to
explore the outer solar system and to send back to Earth their navigational
signals for almost thirty years (for a review see \cite{Turyshev:2010yf} and
references therein).

In recent years, the orbits of Pioneer 10 and 11 were reconstructed very
accurately, by using the original radio-metric Doppler tracking data, based on
the signals exchanged between the spacecraft and NASA's terrestrial tracking
stations. This reconstruction yielded a persistent discrepancy between the
observed and predicted data, equivalent to an unexplained small acceleration
of the spacecraft in the direction of the Sun. This effect is evidenced by
measuring a small frequency shift (toward higher frequencies, i.e., a
\textquotedblleft blueshift\textquotedblright) of the signal reaching us from
the spacecraft. The nature of this anomalous acceleration or of the related
blueshift remains unexplained; this effect has become known as the
\textquotedblleft Pioneer anomaly\textquotedblright\ (\cite{Anderson:1998jd},
\cite{Turyshev:1999tj}, \cite{Anderson:2001sg}).

The importance of this effect, as well as of other known gravitational
anomalies (\cite{Lammerzahl:2006ex}, \cite{Anderson:2009si},
\cite{Turyshev:2009mt}), is not related to how they affect the spacecraft
navigation, since they all produce very small corrections to the orbits, but
to the possibility that these anomalies might be an indication of new
gravitational physics. In particular, several non-conventional explanations of
these effects have been proposed (see discussion in \cite{Turyshev:2010yf},
\cite{Anderson:1998jd}, \cite{Anderson:2001sg}) such as modifications of the
law of gravity, or a modified inertia, as proposed by the Modified Newtonian
Dynamics (MOND) theory, the existence of a dark matter halo\ around the Earth,
or in the solar system, which might slightly alter the gravitational force
acting on the spacecraft, and several others.

In this line of reasoning, alternative gravitational theories such as
Conformal Gravity (CG), originally proposed by H. Weyl in 1918
(\cite{Weyl:1918aa}, \cite{Weyl:1918ib}, \cite{Weyl:1919fi}) and revisited by
P. Mannheim (\cite{Mannheim:1988dj}, \cite{Kazanas:1988qa},
\cite{Mannheim:2005bf}), have provided a new framework for cosmological
models, with the advantage of avoiding some of the most controversial elements
of current standard cosmology, such as dark matter, dark energy, inflation,
and others.

Following the original CG, we have recently studied an alternative approach to
these models which was named \textquotedblleft kinematical conformal
cosmology\textquotedblright\ \cite{Varieschi:2008fc}, but that for brevity
will be called Conformal Cosmology (CC) in the rest of this paper. This
approach was based on the direct application of the conformal symmetry to the
Universe, i.e., considering the possibility that a \textquotedblleft
stretching\textquotedblright\ of the spacetime fabric might be acting over
cosmological scales. In a second part of this work \cite{Varieschi:2008va} it
was shown that this model can successfully fit type-Ia Supernovae data,
without assuming the existence of dark matter or dark energy.

In addition, a preliminary analysis also performed in our second paper
\cite{Varieschi:2008va} indicated that CC might be able to explain the
existence of the Pioneer anomaly, since the observed blueshift of the
spacecraft signal could be due to a region of cosmological blueshift
surrounding our solar system, which is naturally predicted by our model. A new
comprehensive review of the Pioneer anomaly has recently been published
\cite{Turyshev:2010yf}, together with more details of the Pioneer early data
\cite{Nieto:2005kb}, thus prompting us to reconsider and improve our previous
analysis \cite{Varieschi:2008va}, based on the conformal cosmology approach.

In the next section we will briefly review our CC solutions, showing how a
local blueshift region can naturally emerge, while in Sect.
\ref{sect:pioneer_anomaly} we will fit all current Pioneer data
\cite{Nieto:2005kb} with our cosmological solutions and determine the values
of the parameters in our model. Finally, in Sect.
\ref{sect:discussion_conclusions}, we will discuss our results and compare
them to the existing physical limits of standard gravity in the solar system.

\section{\label{sect:conformal_cosmology}Conformal cosmology}

In our first CC paper \cite{Varieschi:2008fc}\ we used as a starting point the
line element originally derived by Mannheim-Kazanas \cite{Mannheim:1988dj} as
an exterior solution for a static, spherically symmetric source in conformal
gravity theory, i.e., the analogue of the Schwarzschild exterior solution in
general relativity:%

\begin{equation}
ds^{2}=-B(r)\ c^{2}dt^{2}+\frac{dr^{2}}{B(r)}+r^{2}d\psi^{2} \label{eqn2.1}%
\end{equation}
where $d\psi^{2}=d\theta^{2}+\sin^{2}\theta\ d\phi^{2}$ in spherical
coordinates and%

\begin{equation}
B(r)=1-\frac{\beta(2-3\beta\gamma)}{r}-3\beta\gamma+\gamma r-\kappa r^{2},
\label{eqn2.2}%
\end{equation}
with the parameters $\beta=\frac{MG}{c^{2}}\ (%
\operatorname{cm}%
)$, $\gamma\ (%
\operatorname{cm}%
^{-1})$, $\kappa\ (%
\operatorname{cm}%
^{-2})$, where $M$ is the mass of the (spherically symmetric) source and $G$
is the gravitational constant. Conformal gravity introduces two new parameters
$\gamma$ and $\kappa$ which are not present in standard general relativity,
while the familiar Schwarzschild solution is recovered in the limit for
$\gamma,\kappa\rightarrow0$, in the equations above.

We then considered regions far away from matter distributions, thus ignoring
the matter dependent $\beta$ terms, and rewrote the last equation in a
simplified form:%

\begin{equation}
B(r)=1+\gamma r-\kappa r^{2}=1+\gamma r+\left(  \frac{\gamma^{2}}{4}+k\right)
r^{2}=-g_{00}(r), \label{eqn2.3}%
\end{equation}
where the parameter $k$ is linked to $\gamma$ and $\kappa$, through
$k=-\frac{\gamma^{2}}{4}-\kappa$ and it is ultimately connected to the
so-called trichotomy constant $\mathbf{k}$ (in bold)\ of a Robertson-Walker
(RW) metric, defined as $\mathbf{k}\equiv\frac{k}{\left\vert k\right\vert
}=0,\pm1$. This is also related to another fundamental aspect of CG: the
existence of coordinate and conformal transformations connecting the static,
spherically symmetric solution represented by Eqs. (\ref{eqn2.1}) and
(\ref{eqn2.3}), with the classical Robertson-Walker metric (see details in
\cite{Varieschi:2008fc}).

It was precisely this connection between the two solutions which prompted us
to consider the CG\ static, spherically symmetric solution as an alternative
description of the standard cosmological evolution, based on the RW metric. In
other words, the CG static solution might also contain information about the
cosmological redshift, the expansion of the Universe, etc., and constitute an
alternative approach to cosmology. In particular, the CG expressions in Eqs.
(\ref{eqn2.2}) or (\ref{eqn2.3}) contain a linear and a quadratic term, in the
radial coordinate $r$, which might yield considerable effects\footnote{By
using the contribution of the linear term $\gamma r$, the flat galactic
rotation curves were in fact explained by Mannheim (\cite{Mannheim:1992vj},
\cite{Mannheim:1996rv}) without the need of dark matter.} at large distances,
including a strong gravitational redshift which could be, at least in part,
responsible for the observed cosmological redshift.

Therefore, we postulated in Ref. \cite{Varieschi:2008fc} that the observed
redshift is due to this gravitational\ effect, which influences the wavelength
or frequency of a light signal emitted at time $t$ and position $r$, and
observed at the origin ($r=0$) at the current time $t_{0}$, in the following way:%

\begin{equation}
1+z=\frac{R(0)}{R(r)}=\frac{\lambda(r,t)}{\lambda(0,t_{0})}=\frac{\nu
(0,t_{0})}{\nu(r,t)}=\sqrt{\frac{-g_{00}(0)}{-g_{00}(r)}}=\frac{1}%
{\sqrt{1+\gamma r+\left(  \frac{\gamma^{2}}{4}+k\right)  r^{2}}}.
\label{eqn2.4}%
\end{equation}

In the previous equation the redshift factor $(1+z)$ is related to the ratio
of cosmic scale factors $R$, which simply depend on the radial distance $r$,
in view of Eq. (\ref{eqn2.3}). Alternatively, to obtain a time dependent form
of the cosmic scale factor, we considered that the radial distance $r$ is
associated with a look-back time $(t_{0}-t)$, related to the time of travel of
a light signal. Integrating the CG metric in Eqs. (\ref{eqn2.1}) and
(\ref{eqn2.3}) along the null geodesic, we obtained \cite{Varieschi:2008fc}:%

\begin{align}
1+z  &  =\frac{R(t_{0})}{R(t)}=\left(  \cos\chi-\delta\sin\chi\right)
\ ;\ k>0\label{eqn2.5}\\
1+z  &  =\frac{R(t_{0})}{R(t)}=\left(  1-\delta\chi\right)
\ ;\ k=0\nonumber\\
1+z  &  =\frac{R(t_{0})}{R(t)}=\left(  \cosh\chi-\delta\sinh\chi\right)
\ ;\ k<0,\nonumber
\end{align}
for the three possible values of the parameter $k$. In the previous equation
we preferred to use dimensionless quantities and parameters, defined as follows:%

\begin{equation}
\chi\equiv\sqrt{\left\vert k\right\vert }c(t_{0}-t)\ ;\ \delta\equiv
\frac{\gamma}{2\sqrt{\left\vert k\right\vert }}, \label{eqn2.6}%
\end{equation}
so that the fundamental parameters of our conformal cosmology are now
expressed by $\gamma$ $(%
\operatorname{cm}%
^{-1})$ and the dimensionless $\delta$ ($c$ is the speed of light in vacuum,
assumed constant).

\begin{figure}[ptb]%
\centering
\fbox{\ifcase\msipdfoutput
\includegraphics[
width=\textwidth
]
{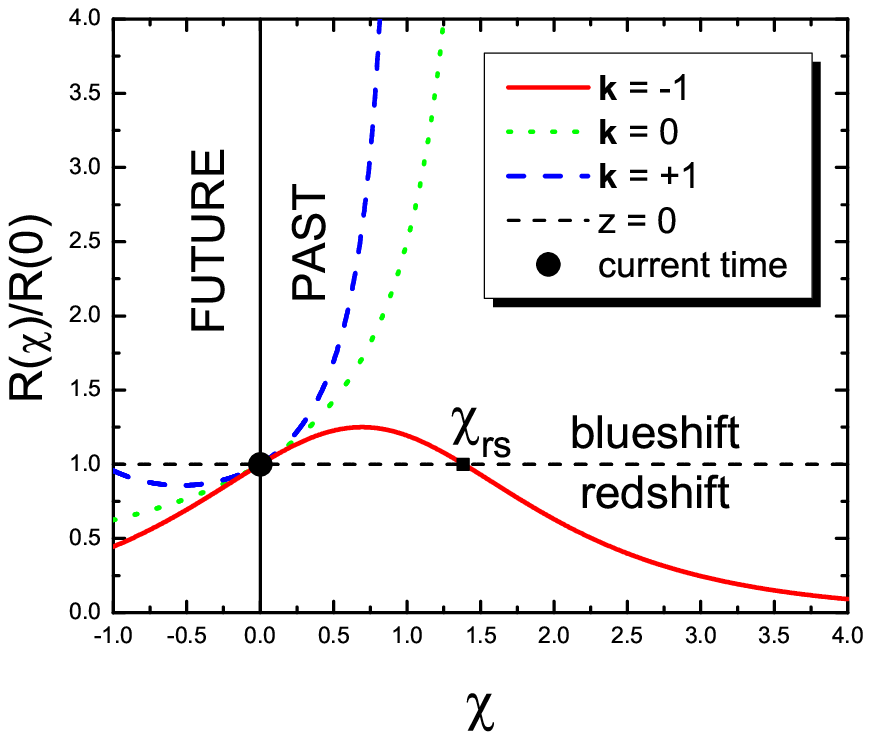}%
\else
\includegraphics[
height=5.3033in,
width=6.3023in
]%
{C:/swp55/Docs/KINEMATICAL3/arXiv/graphics/Figure1__1.pdf}%
\fi
}\caption[$R$ functions obtained from Eq. (\ref{eqn2.5}) are shown here for
different values of $\mathbf{k}$.]{$R$ functions obtained from Eq.
(\ref{eqn2.5}) are shown here for different values of $\mathbf{k}$:
$\mathbf{k}=-1$ in red (solid), $\mathbf{k}=0$ in green (dotted), and
$\mathbf{k}=+1$ in blue (dashed), and for a positive value of the parameter
$\delta\simeq0.6$ (an unrealistically large value; our current value
$\delta=\delta(t_{0})$ will be shown to be positive and close to zero.}%
\label{fig1}%
\end{figure}

In Fig. \ref{fig1} we plot the results of Eq. (\ref{eqn2.5}) in terms of the
inverse ratio $R(\chi)/R(0)=R(t)/R(t_{0})=1/(1+z)$ which describes better the
cosmic evolution. The dimensionless quantity $\chi\equiv\sqrt{\left\vert
k\right\vert }c(t_{0}-t)$, on the horizontal axis, represents a look-back
time, so that the universal evolution of the cosmic scale factor, from the
past to the future, can be seen by following our curves from right to left.
The circular dot on the vertical axis represents our \textquotedblleft current
time\textquotedblright\ ($\chi=0$). We can clearly see that the only solution
which shows a redshift in the past (values below the horizontal black dashed
line, representing $z=0$) is the red-solid curve, corresponding to
$\mathbf{k}=-1$. Therefore, the other two solutions, for $\mathbf{k}=+1$, $0$,
are ruled out; only the $\mathbf{k}=-1$ solution will be considered in the following.%

Our preferred solution in Fig. \ref{fig1} (red-solid) also shows a blueshift
region in the immediate past of our current time, which in Sect.
\ref{sect:pioneer_anomaly} will be related directly to the Pioneer anomaly.
This blueshift region is greatly exaggerated in the figure, since the
different curves were plotted for $\delta\simeq0.6$, an unrealistically high
value. We will show in the next sections that $\delta$ is positive and close
to zero, resulting in a very small-sized blueshift region, compared to the
overall size of the Universe. Similar plots can be obtained for the ratio
$R/R_{0}$ expressed in terms of the radial distance $r$ (see
\cite{Varieschi:2008fc}\ for details), which also suggest the existence of a
blueshift region localized around the observer's position, i.e., the Earth
could be surrounded by a natural blueshift region, extending at least over the
solar system region. This might be the origin of the Pioneer
anomaly.\footnote{Obviously, the Earth's observer is not located at any
privileged position. The same cosmological evolution described by CC would be
seen by any other observer in the Universe, provided that the local values of
the cosmological parameters $\delta$ and $\gamma$ are the same. In our
previous work (\cite{Varieschi:2008fc}, \cite{Varieschi:2008va}) we have
suggested that $\delta$ might play the role of a universal time, so that for a
certain value of this parameter the evolution of the Universe would look the
same for any observer. In this way, conformal cosmology does not violate the
cosmological principle, which postulates a homogeneous and isotropic
Universe.}

Before we proceed to analyze this possible explanation for the anomaly, we
recall a few more results obtained in our second paper \cite{Varieschi:2008va}%
. Since we have closed-form expressions for our scale factor $R$, in Eqs.
(\ref{eqn2.4}) and (\ref{eqn2.5}), it is straightforward to obtain the Hubble
parameter ($H(t)=\overset{\cdot}{R}(t)/R(t)$) and the deceleration parameter
($q(t)=-\frac{\overset{\cdot\cdot}{R}(t)}{R(t)H^{2}(t)}=-\frac{\overset{\cdot
\cdot}{R}(t)R(t)}{\overset{\cdot}{R}{}^{2}(t)}$) as a function of time or
redshift $z$. For the $\mathbf{k}=-1$ case, we obtained
\cite{Varieschi:2008va}:%

\begin{align}
H(t)  &  =\sqrt{\left\vert k\right\vert }c\left(  \frac{\sinh\chi-\delta
\cosh\chi}{\cosh\chi-\delta\sinh\chi}\right)  =\pm\sqrt{\left\vert
k\right\vert }c\frac{\sqrt{(1+z)^{2}-(1-\delta^{2})}}{(1+z)}\label{eqn2.7}\\
q(t)  &  =\left(  \frac{\cosh\chi-\delta\sinh\chi}{\sinh\chi-\delta\cosh\chi
}\right)  ^{2}-2=\frac{(1+z)^{2}}{(1+z)^{2}-(1-\delta^{2})}-2,\nonumber
\end{align}
and, in particular for $\chi\rightarrow0$ or $z\rightarrow0$:%

\begin{align}
H(t_{0})  &  =-\frac{\gamma}{2}c\ ;\ H(z=0)=\pm\frac{\gamma}{2}c\label{eqn2.8}%
\\
q(t_{0})  &  =q(z=0)=\frac{1}{\delta^{2}}-2.\nonumber
\end{align}

The signs of the quantities in Eqs. (\ref{eqn2.7}) and (\ref{eqn2.8}) can be
explained by considering again the red-solid curve in Fig. \ref{fig1}, which
represents the ratio $R(\chi)/R(0)$, or equivalently $R(t)/R(t_{0})$, over
different cosmological epochs. This bell-shaped curve was plotted for a
positive value of $\delta$ and shows a local blueshift area in the
\textquotedblleft past\textquotedblright\ evolution of the Universe, extending
back to $\chi_{rs}=\operatorname{arccosh}\left[  (1+\delta^{2})/(1-\delta
^{2})\right]  =2\operatorname{arctanh}\delta$ (represented by the square point
in Fig. \ref{fig1}) or $(t_{0}-t_{rs})=\frac{2}{\sqrt{\left\vert k\right\vert
}c}\operatorname{arctanh}\delta$, for the look-back time at which the redshift
($rs$) starts being observed. The red curve has a maximum at $\chi_{\max
}=\operatorname{arctanh}\delta$ or $(t_{0}-t)_{\max}=\frac{1}{\sqrt{\left\vert
k\right\vert }c}\operatorname{arctanh}\delta$ (we can also find $(R(\chi
)/R(0))_{\max}=1/\sqrt{1-\delta^{2}}$ or $z_{\min}=\sqrt{1-\delta^{2}}-1$) and
it is evidently symmetric around this point of maximum expansion of the Universe.

Therefore, for each value of $z$, i.e., for each value of $R(\chi)/R(0)$, we
have two corresponding values of the Hubble parameter (except at the maximum,
for $z_{\min}=\sqrt{1-\delta^{2}}-1$, where $H=0$). The two related points on
the curve, at the same redshift level, will have equal and opposite expansion
rates. This yields the double sign in the previous expressions for $H$, when
given as a function of $z$. This argument applies also to the $z=0$ case,
corresponding to the current time $t_{0}$, at which $H(t_{0})=-\frac{\gamma
}{2}c$ is negative, showing that the Universe is already in a contracting
phase.\footnote{This is also a consequence of the signs of our conformal
parameters, in particular the positive value of $\gamma$. Our estimate of
$\gamma$ will be given in Sect. \ref{sect:pioneer_anomaly}, but we recall that
Mannheim has independently evaluated $\gamma$ as a small but positive quantity
($\gamma_{Mann}=3.06\times10^{-30}\ cm^{-1}$), by fitting rotational velocity
curves for several spiral galaxies, using conformal gravity
\cite{Mannheim:2005bf}. If $\gamma$ were to have a negative value, we would
still be in an expanding phase of the Universe.} As discussed above, the same
$z=0$ value can also refer to the time in the past ($t_{rs}$) at which we
start observing the cosmological redshift, with $H(t_{rs})=+\frac{\gamma}{2}%
c$, a positive quantity. This analysis does not contradict the current
astrophysical estimates of $H_{0}$ as a positive quantity. They are based on
redshift observations of light coming from galaxies at times in the past
$t\lesssim t_{rs}$, therefore what is denoted by $H_{0}$ in standard cosmology
should be actually indicated as $H(t_{rs})=+\frac{\gamma}{2}c$, again a
positive quantity related to the expanding phase of the Universe. The same
analysis can be done in terms of radial distances $r$. The blueshift region
would extend from $r=0$ up to a distance given by:
\begin{equation}
r_{rs}=\gamma/(\left\vert k\right\vert -\frac{\gamma^{2}}{4})=\frac{4}{\gamma
}\frac{\delta^{2}}{1-\delta^{2}}, \label{eqn2.8.1}%
\end{equation}
where $r_{rs}$\ is the distance at which we start observing the cosmological
redshift. In general, the slope of the red-solid curve in Fig. \ref{fig1} is
related to the value of the Hubble parameter at that point, while its
curvature is connected to the deceleration parameter, through the expressions
given above.

In particular, following Eq. (\ref{eqn2.8}), the slope of the plot and its
curvature at current time $t_{0}$ are basically connected to our two
fundamental parameters $\gamma$ and $\delta$. In the next section we will show
that the slope of the red-solid plot at $t_{0}$ is closely related to the
value of the Pioneer anomalous acceleration $a_{P}$, which can therefore be
used to determine $\gamma$. Similarly, the curvature of the plot at $t_{0}$
will be related to the rate of change of the anomalous acceleration (i.e., the
\textquotedblleft jerk\textquotedblright\ $j_{P}\equiv\overset{\cdot}{a}_{P}$)
and will be used to determine the value of our other parameter $\delta$.

We conclude this section by noting that the values of our parameters ($\delta$
and $\gamma$) could be derived directly from standard cosmological
observations, in view of Eq. (\ref{eqn2.8}). Using the current best estimate
of $H_{0}=(72\pm3)\ km\ s^{-1}\ Mpc^{-1}$ \cite{Nakamura:2010} and the
positive sign in Eq. (\ref{eqn2.8}) we obtain:%

\begin{equation}
\gamma=\frac{2H_{0}}{c}=(1.56\pm0.06)\times10^{-28}\ cm^{-1}. \label{eqn2.9}%
\end{equation}
The direct determination of $\delta$ is more difficult, since the deceleration
parameter $q$ is not known explicitly. In Ref. \cite{Varieschi:2008va} we
based our analysis on recent luminosity data for type-Ia Supernovae, obtaining
an estimate of $\delta\simeq3.83\times10^{-5}$, but this analysis needs to be
confirmed by further studies.

\section{\label{sect:pioneer_anomaly}The Pioneer anomaly}

In the previous section we briefly reviewed our conformal cosmology and
outlined the reasons why we consider the $\mathbf{k}=-1$ solution as a
possible description of the evolution of the Universe. This solution can
explain the observed cosmological redshift, but it requires the existence of a
blueshift region in the immediate vicinity of our current spacetime position
in the Universe.

This could be a serious problem for our model, since we do not observe
blueshift of nearby astrophysical objects except for the one caused by the
peculiar velocities of nearby galaxies, presumably due to standard Doppler
shift. However, as already mentioned in Sect. \ref{sect:introduction},
experimental evidence of a local region of blueshift might come from the
analysis of the Pioneer anomaly (\cite{Turyshev:2010yf},
\cite{Anderson:1998jd}, \cite{Turyshev:1999tj}, \cite{Anderson:2001sg},
\cite{Nieto:2005kb}, \cite{Anderson:2001ks}, \cite{Turyshev:2005zk},
\cite{Turyshev:2005ej}, \cite{Toth:2006qb}, \cite{Toth:2007uu},
\cite{Nieto:2007tf}, \cite{Toth:2009se}, \cite{Turyshev:2009hh}).

This is a small frequency drift (blueshift), observed analyzing the
navigational data of the Pioneer 10-11 spacecraft, received from distances
between $20-70\ AU$ (astronomical units) from the Sun, while these spacecraft
were exploring the outer solar system. This anomaly is usually reported as a
positive rate of change of the signal frequency, $\overset{\cdot}{\nu}_{P}>0$
(blueshift), resulting in a frequency drift of about $1.5\ Hz$ every $8$
years, or as an almost constant sunward acceleration, $a_{P}<0$, or even as a
\textquotedblleft clock acceleration\textquotedblright\ $a_{t}\equiv
\frac{a_{P}}{c}<0$. More precisely (\cite{Turyshev:2010yf},
\cite{Anderson:2001ks}):%

\begin{align}
\overset{\cdot}{\nu}_{P}  &  =(5.99\pm0.01)\times10^{-9}\ s^{-2}%
\label{eqn3.1}\\
a_{P}  &  =-(8.74\pm1.33)\times10^{-8}\ cm\ s^{-2}\nonumber\\
a_{t}  &  \equiv\frac{a_{P}}{c}=-(2.92\pm0.44)\times10^{-18}\ s^{-1}.\nonumber
\end{align}

An attempt was made to detect such anomaly also in the radiometric data from
other spacecraft traveling at the outskirts of the solar system, such as the
Galileo and Ulysses missions \cite{Anderson:2001ks}. In the case of Galileo,
the effects of solar radiation made such detection impossible, while for
Ulysses a possible anomalous acceleration $a_{Ulysses}=-(12\pm3)\times
10^{-8}\ cm/s^{2}$ was seen in the data. Other spacecraft, such as the New
Horizons mission to Pluto, launched in 2006, might provide new data in the
near future. These discoveries prompted a complete re-analysis of all the
historical navigational data of these space missions, which is currently
underway (\cite{Turyshev:2010yf}, \cite{Turyshev:2005ej}, \cite{Toth:2006qb},
\cite{Toth:2007uu}, \cite{Turyshev:2009hh}, \cite{Turyshev:2007ii}) and will
be completed in the next few months \cite{Slava:1}. This new analysis will try
to determine additional characteristics of the anomaly, such as its precise
direction, the possible temporal and spatial variations, its dependence on
heliocentric or geocentric distance, etc. A future dedicated mission is also
being proposed (\cite{Anderson:2002yc}, \cite{Nieto:2004np},
\cite{Dittus:2005re}, \cite{Turyshev:2005zm}) to test directly this puzzling phenomenon.

Currently, the origin and nature of this anomaly remains unexplained; all
possible sources of systematic errors have been considered
(\cite{Turyshev:2010yf}, \cite{Anderson:2001sg}, \cite{Anderson:2001ks},
\cite{Turyshev:2005zk}, \cite{Toth:2007uu}, \cite{Turyshev:2007ii},
\cite{Nieto:2003rq}) but they cannot fully account for the observed effect.
The current focus of conventional explanations of the anomaly seems to be the
thermal recoil force, i.e., anisotropically emitted thermal radiation,
originating from the spacecraft four radioisotope thermoelectric generators
(RTGs), which can contribute significantly to the measured acceleration. The
natural decay of the radioactive material in the RTGs, the aging of the
thermocouples in the system and other effects, all contribute to the decrease
of the total thermal power during the spacecraft life. This might explain the
decrease over time of the measured Pioneer acceleration (in absolute value),
i.e., the negative \textquotedblleft jerk\textquotedblright\ $\left\vert
\overset{\cdot}{a}_{P}\right\vert <0$, already seen in the early Pioneer data
(\cite{Turyshev:2010yf}, \cite{Nieto:2005kb}, \cite{Slava:1}).

Although the anomaly can be caused by standard physical effects, we will try
in the following to explain its origin by using the cosmological model
outlined in the previous section. The phenomenology of the Pioneer anomaly is
related to a complex exchange of radiometric signals between the tracking
stations on Earth (of the Deep Space Network - DSN) and the spacecraft, using
S-band Doppler frequencies ($1.55-5.20\ GHz$). Typically, an uplink signal is
sent from the DSN to the spacecraft at a frequency of $2.11\ GHz$, based on a
very stable hydrogen maser system, then an S-band transponder onboard the
spacecraft applies an exact and fixed turn-around ratio of $240/221$ to the
uplink signal, so that the Pioneer returns a downlink signal at a slightly
different frequency of about $2.29\ GHz$, to avoid interference with the
uplink one.

This procedure is known as a two-way Doppler coherent mode and allows for very
precise tracking of the spacecraft, since the returning signal is directly
compared to the original one. On the contrary, a one-way Doppler signal (with
a fixed signal source on the spacecraft, whose frequency cannot be monitored
for accuracy) is less effective. This type of tracking system added to the
propulsion and navigational characteristics of the Pioneer spaceship
(especially the presence of a spin-stabilization system) resulted in a very
good acceleration sensitivity of about $10^{-8}\ cm/s^{2}$, once the influence
of solar radiation pressure can be neglected (for distances $\gtrsim20\ AU$
from the Sun).

The DSN station acquires the downlink signal after a time delay ranging from a
few minutes to some hours, depending on the distance involved, and compares it
to the reference frequency to determine the Doppler shift due to the actual
motion of the spacecraft. The navigational software can also model with great
precision the expected frequency of the signal returned from the Pioneer,
which should coincide with the one observed on Earth. As already mentioned, a
discrepancy was found, corresponding to the values in Eq. (\ref{eqn3.1}),
whose origin cannot be traced to any systematic effect due to either the
performance of the spacecraft or the theoretical modeling of its navigation.

The Pioneer anomaly was first reported (\cite{Anderson:1998jd},
\cite{Turyshev:1999tj}, \cite{Anderson:2001sg}) as an almost constant value of
the anomalous acceleration, with temporal and space variation of $a_{P}$
within $10\%$, over a range of heliocentric distances $\sim20-70\ AU$, and
possibly at even closer distances $\lesssim10\ AU$, so that we will
concentrate first on the average value of $a_{P}$ and later on its variation
with time and distance. In our view, the Pioneer phenomenology represents the
most basic experiment we could perform in order to check if the cosmic
evolution is really affecting the frequency of electromagnetic radiation
emitted and observed at different spacetime locations, following Eqs.
(\ref{eqn2.4}) and (\ref{eqn2.5}).

In the standard analysis of the Pioneer anomaly, the signal coming back to
Earth is affected by the relativistic Doppler effect. Following this model,
$\nu_{\operatorname{mod}}$ will be the frequency of the expected signal and
will be related to the signal reference frequency $\nu_{ref}$ $=2.11\ GHz$
(for the uplink signal in a two-way system) by the standard relativistic
Doppler formula (see Eq. 2.2.2 in \cite{Weinberg}):%

\begin{equation}
\frac{\nu_{\operatorname{mod}}}{\nu_{ref}}=\frac{\sqrt{1-\frac{\mathbf{v}^{2}%
}{c^{2}}}}{1+\frac{v_{r}}{c}}\simeq1-\frac{v_{r}}{c}, \label{eqn3.2}%
\end{equation}
where $v_{r}$ is the spacecraft radial velocity and the approximation on the
right-hand side holds to first order in $v_{r}/c$.

Since we have a two-way system, the Doppler shift involved is actually double,
so we can use the previous equation but with $v_{r}=2v_{\operatorname{mod}%
}(t^{\prime})$, where $v_{\operatorname{mod}}(t^{\prime})$ is the expected
velocity of the spacecraft, according to the theoretical navigation model, at
time $t^{\prime}$, when the spaceship receives and immediately re-transmits
the signal. We use here a time variable $t^{\prime}$ which can be simply
considered the elapsed time since the spacecraft launch ($t^{\prime}=0$ at
$r=0$) and then later we will simply identify $t^{\prime}$ with our
cosmological look-back time $(t_{0}-t)$ in Eq. (\ref{eqn2.6}). With this
radial velocity, Eq. (\ref{eqn3.2}) to first order in $v_{r}/c$ becomes:%

\begin{equation}
\nu_{\operatorname{mod}}(t^{\prime})\simeq\nu_{ref}\left[  1-\frac
{2v_{\operatorname{mod}}(t^{\prime})}{c}\right]  \label{eqn3.3}%
\end{equation}
and this frequency is expected to be observed with high precision, due to the
reported excellent navigational control of the spacecraft.

On the contrary, a different frequency is observed, $\nu_{obs}(t^{\prime}%
)>\nu_{\operatorname{mod}}(t^{\prime})$, involving an additional unexplained
blueshift: this is the Pioneer anomaly. Following Eq. (\ref{eqn3.1}), the
frequency difference is reported as:%

\begin{align}
\Delta\nu(t^{\prime})  &  =\nu_{obs}(t^{\prime})-\nu_{\operatorname{mod}%
}(t^{\prime})\simeq2t^{\prime}\overset{\cdot}{\nu}_{P}\label{eqn3.4}\\
\overset{\cdot}{\nu}_{P}  &  =5.99\times10^{-9}\ s^{-2}\text{ (one-way)}%
\nonumber
\end{align}
where the factor of two in the first line of the previous equation is due to
the two-way system. We also remark here that several of the cited references
adopt a rather confusing \textquotedblleft DSN sign
convention\textquotedblright\ for the frequency difference in Eq.
(\ref{eqn3.4}) (see \cite{Turyshev:2010yf}, \cite{Anderson:2001sg},
\cite{Toth:2006qb} and Ref. (38) of \cite{Anderson:1998jd}), resulting in a
change of sign in most of their equations. We prefer to use here our
definition of $\Delta\nu$ as given in the previous equation.

The anomalous acceleration $a_{P}$ is introduced as an alternative way of
describing the effect, although in our view it does not correspond to a real
spacecraft acceleration. As in Eq. (\ref{eqn3.3}), we can write the observed
frequency to first order in $v_{r}/c$ as:%

\begin{equation}
\nu_{obs}(t^{\prime})\simeq\nu_{ref}\left[  1-\frac{2v_{obs}(t^{\prime})}%
{c}\right]  , \label{eqn3.5}%
\end{equation}
where the \textquotedblleft observed\textquotedblright\ velocity of the
spacecraft refers to the time of interest $t^{\prime}$. Combining together the
last three equations we can write the frequency difference as:%

\begin{equation}
\Delta\nu(t^{\prime})=-2\frac{\nu_{ref}}{c}\left[  v_{obs}(t^{\prime
})-v_{\operatorname{mod}}(t^{\prime})\right]  =-2\frac{\nu_{ref}}{c}\Delta
v(t^{\prime}). \label{eqn3.6}%
\end{equation}

These frequency differences $\Delta\nu$ (also called frequency residuals in
the literature cited) are therefore equivalent to the corresponding velocity
residuals ($\Delta v=v_{obs}-v_{\operatorname{mod}}$) and they are usually
plotted as a function of the elapsed time $t^{\prime}$, showing an almost
linear increase with time of these residuals, which is the essence of the
Pioneer anomaly (see for example Fig. 5.2\ in Ref. \cite{Turyshev:2010yf}).
The Pioneer anomalous acceleration can be defined as the rate of change of the
velocity residuals, related to the corresponding rate of change of the
frequency residuals, in view of Eq. (\ref{eqn3.6}). Therefore, if we define
$a_{P}\equiv\frac{d\left(  \Delta v\right)  }{dt^{\prime}}\simeq\left[  \Delta
v(t^{\prime}+\Delta t^{\prime})-\Delta v(t^{\prime})\right]  /\Delta
t^{\prime}$, the Pioneer acceleration can be related to the frequency differences%

\begin{equation}
a_{P}=-\frac{c}{2\nu_{ref}}\frac{d\left(  \Delta\nu\right)  }{dt^{\prime}%
}\simeq-\frac{c}{2\nu_{ref}}\frac{\Delta\nu(t^{\prime}+\Delta t^{\prime
})-\Delta\nu(t^{\prime})}{\Delta t^{\prime}}, \label{eqn3.7}%
\end{equation}
which are more significant quantities in our analysis. We will assume that
these frequency differences are intrinsically due to the different locations
of the spacecraft (at position $r$) and of the Earth's observer (at $r=0$).
Therefore, we identify the reference frequency $\nu_{ref}$ in Eq.
(\ref{eqn3.3}) with $\nu(0)$ and the similar quantity $\nu_{ref}$ in Eq.
(\ref{eqn3.5}) with $\nu(r)$. Then, we subtract Eq. (\ref{eqn3.3}) from Eq.
(\ref{eqn3.5}):%

\begin{align}
\Delta\nu(t^{\prime})  &  =2\left[  \nu(r)-\nu(0)\right]  \left[
1-\frac{2v(t^{\prime})}{c}\right]  \simeq2\left[  \nu(r)-\nu(0)\right]
=2\nu(0)\left[  \frac{\nu(r)}{\nu(0)}-1\right]  =\label{eqn3.8}\\
&  =2\nu_{ref}\left[  \frac{R(r)}{R(0)}-1\right]  =2\nu_{ref}\left[
\frac{R(t^{\prime})}{R(0)}-1\right]  \simeq2\nu_{ref}\left[  \frac{\gamma}%
{2}t^{\prime}\right]  ,\nonumber
\end{align}
where the common factor of two in all the parts of the previous equation was
added again because of the two-way effect, which has to be included also in
our gravitational blueshift model. The velocities $v_{\operatorname{mod}%
}(t^{\prime})$ and $v_{obs}(t^{\prime})$ from Eqs. (\ref{eqn3.3}) and
(\ref{eqn3.5}) are assumed to be the same, so that the common factor $\left[
1-\frac{2v(t^{\prime})}{c}\right]  \simeq1$ is close to unity and can be
neglected, since the average Pioneer speed is $v_{P}\simeq12.8\ km/s\ll c$
\cite{Markwardt:2002ma}.\ We also identified $\nu(0)$ with the Earth reference
frequency $\nu_{ref}$ and used our fundamental Eq. (\ref{eqn2.4}) and Eq.
(\ref{eqn2.5}), $k<0$ case, to first order in $\chi=\sqrt{\left\vert
k\right\vert }ct^{\prime}$.\footnote{The elapsed time $t^{\prime}$ for the
Pioneer spacecraft missions is of the order of a few years ($1yr=3.156\times
10^{7}s$); we can assume $\sqrt{\left\vert k\right\vert }\sim\gamma
\sim10^{-28}-10^{-30}\ cm^{-1}$, therefore $\sqrt{\left\vert k\right\vert
}ct^{\prime}\sim10^{-10}-10^{-12}\ll1$.} Similarly, we have: $\Delta
\nu(t^{\prime}+\Delta t^{\prime})\simeq2\nu_{ref}\left[  \frac{\gamma}%
{2}\left(  t^{\prime}+\Delta t^{\prime}\right)  \right]  $ so that Eq.
(\ref{eqn3.7}) simplifies as follows:%

\begin{equation}
a_{P}=a_{P}(t_{0})=-\frac{\gamma}{2}c^{2}=cH(t_{0}), \label{eqn3.9}%
\end{equation}
in view also of our evaluation of $H(t_{0})=-\frac{\gamma}{2}c$ (a negative
quantity) from Eq. (\ref{eqn2.8}).

This result immediately explains the often cited \textquotedblleft numerical
coincidence,\textquotedblright\ i.e., the simple relation $\left\vert
a_{P}\right\vert \simeq cH_{0}$ between the Pioneer acceleration and the
standard (positive) Hubble constant, with the correct negative sign for both
quantities in Eq. (\ref{eqn3.9}), in view of our previous discussion of the
sign of $H(t_{0})<0$. Eq. (\ref{eqn3.9}) can also be used to determine
$\gamma$ and $H_{0}$ (as a positive quantity), using the reported value of
$a_{P}$ from Eq. (\ref{eqn3.1}):%

\begin{align}
\gamma &  =\gamma(t_{0})=-\frac{2}{c^{2}}a_{P}=\left(  1.94\pm0.30\right)
\times10^{-28}\ cm^{-1},\label{eqn3.10}\\
H_{0}  &  =\left(  90.0\pm13.7\right)  \ km\ s^{-1}\ Mpc^{-1}.\nonumber
\end{align}

The value of $\gamma$ (considered measured at the current time $t_{0}$, even
if the Pioneer data are a few years old) is close to our first direct estimate
in Eq. (\ref{eqn2.9}) and the corresponding value of the Hubble constant is
close to the value of standard cosmology. We remark here again that our model
fully explains the reason of this \textquotedblleft numerical
coincidence\textquotedblright\ and provides also the correct signs for all the
quantities involved.\footnote{The numerical \textquotedblleft
coincidence\textquotedblright\ between the Hubble constant and the value of
the Pioneer acceleration $a_{P}$ divided by $c$, was noticed immediately after
the discovery of the Pioneer effect and prompted many speculations and
different explanations. This coincidence is even more striking if one uses the
value cited in Ref. \cite{Anderson:2001sg} as the experimental value for
Pioneer 10 data before systematics, $a_{P}=-7.84\times10^{-8}\ cm\ s^{-2}$,
thus obtaining $H_{0}=80.7\ km\ s^{-1}\ Mpc^{-1}$ and $\gamma_{0}%
=1.74\times10^{-28}\ cm^{-1}$.}

Following Eqs. (\ref{eqn3.7})-(\ref{eqn3.9}) and the related discussion we can
generalize our expression of the Pioneer acceleration, as a function of time
$t^{\prime}$:%

\begin{equation}
a_{P}=-c\frac{d\left[  R(t^{\prime})/R(0)\right]  }{dt^{\prime}}=c^{2}%
\frac{\gamma}{2\delta}\frac{\left[  \sinh\chi-\delta\cosh\chi\right]
}{\left[  \cosh\chi-\delta\sinh\chi\right]  ^{2}}, \label{eqn3.11}%
\end{equation}
with $\chi=\sqrt{\left\vert k\right\vert }ct^{\prime}=\frac{\gamma}{2\delta
}ct^{\prime}$.\footnote{Although $t^{\prime}$ is the elapsed time since the
spacecraft launch, it is treated here as equivalent to a look-back time
$(t_{0}-t)$ because the Pioneer is moving toward increasing distances $r$,
therefore corresponding to increased look-back times in our original redshift
interpretation.} In particular, by using the previous equation and taking
another time derivative, it is easy to derive the \textquotedblleft
jerk\textquotedblright\ $j_{P}\equiv da_{P}/dt^{\prime}$ and its value in the
limit for $t^{\prime}\rightarrow0$:%

\begin{equation}
j_{P}=j_{P}(t_{0})=c^{3}\left(  \frac{\gamma}{2\delta}\right)  ^{2}\left(
1-2\delta^{2}\right)  , \label{eqn3.12}%
\end{equation}
expressed in terms of our fundamental parameters $\gamma$ and $\delta$. The
current value of $j_{P}$ in the last equation is positive (for small values of
$\delta$), but the Pioneer acceleration, as in Eq. (\ref{eqn3.1}) or Eq.
(\ref{eqn3.9}), is considered negative in this paper so that a positive jerk
means that the absolute value of $a_{P}$ will decrease for increasing times or
radial distances, which is indeed shown in the early Pioneer data, as it was
already mentioned at the beginning of this section.%

\begin{figure}[ptb]%
\centering
\fbox{\ifcase\msipdfoutput
\includegraphics[
width=\textwidth
]
{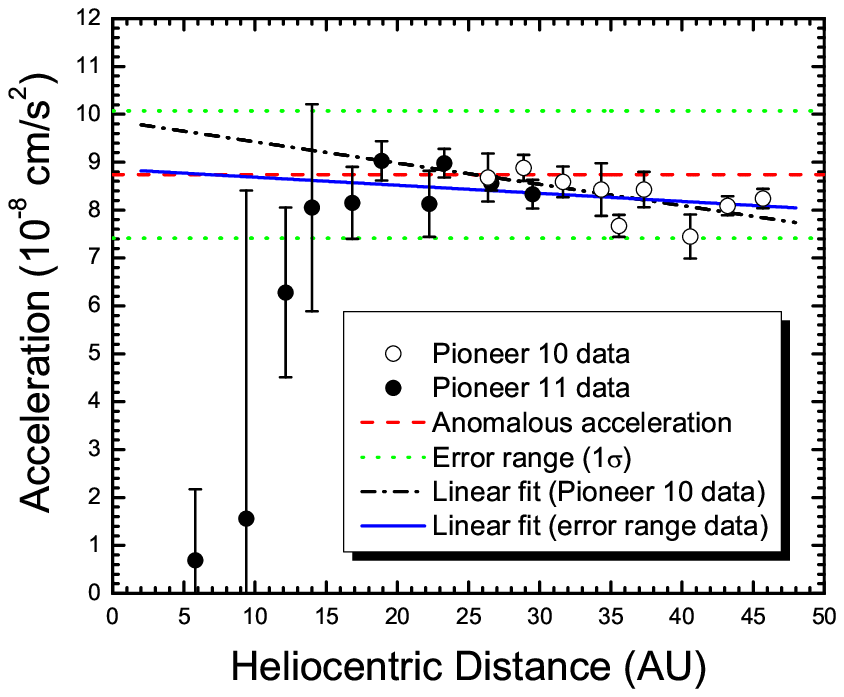}%
\else
\includegraphics[
height=5.152in,
width=6.3023in
]%
{C:/swp55/Docs/KINEMATICAL3/arXiv/graphics/Figure2__2.pdf}%
\fi
}\caption[Early data for Pioneer 10/11 acceleration as a function of
heliocentric distance.]{Early data for Pioneer 10/11 acceleration as a
function of heliocentric distance. The average value of the anomalous
acceleration is indicated in red-dashed, together with its error range
(green-dotted). We also show linear fits of the data, which allow for the
determination of our cosmological parameters $\gamma$ and $\delta$.}%
\label{fig2}%
\end{figure}

In Fig. \ref{fig2} we illustrate the early Pioneer 10/11 data, as originally
reported in Ref. \cite{Nieto:2005kb}, where the absolute value of the Pioneer
acceleration $\left\vert a_{P}\right\vert $ is plotted as a function of the
radial heliocentric distance in AU. The red-dashed horizontal line and the
green-dotted lines represent respectively the value of $\left\vert
a_{P}\right\vert $ and the related 1-sigma error range quoted in Eq.
(\ref{eqn3.1}). The first three data points for Pioneer 11, at smaller
distances, lie outside the considered error range probably because the
anomalous acceleration was masked by solar radiation or other effects. We will
not include these first three data points in our subsequent discussion. We
will concentrate our analysis on either just the Pioneer 10 data points, or
the combination of data points for both spacecraft, but within the 1-sigma
error range (\textquotedblleft error range data\textquotedblright\ in the following).

These two sets of data clearly show a possible decrease of the Pioneer anomaly
(in the absolute value $\left\vert a_{P}\right\vert $) with increasing
heliocentric distance. The black (dash-dotted) line and the blue-solid line in
the same figure represent linear fits for the Pioneer 10 and the error range
data respectively, both of them indicating a decrease of $\left\vert
a_{P}\right\vert $.

If our conformal cosmology is the origin of the Pioneer anomaly, and not the
thermal recoil force mentioned at the beginning of this section, our
\textquotedblleft jerk\textquotedblright\ equation (\ref{eqn3.12}) will
explain the decrease of $\left\vert a_{P}\right\vert $ and can also be used to
determine our second parameter $\delta$.

We computed the slopes of our two linear fits in Fig. \ref{fig2} and used them
as (positive) values of $j_{P}$ in Eq. (\ref{eqn3.12}), together with the
$\gamma$ value from Eq. (\ref{eqn3.10}).\footnote{The radial distances $r$ of
the data plotted in Fig. \ref{fig2} were converted into elapsed times
$t^{\prime}$, by using a simple approximation: $r\simeq v_{P}t^{\prime}$,
where $v_{P}$ is the average Pioneer speed. From the original data (available
from the NASA website at: http://cohoweb.gsfc.nasa.gov/helios/) we estimated:
$v_{P10}\simeq12.96\ km/s$, $v_{P11}\simeq11.42\ km/s$, and used an average
$v_{P}\simeq12.19\ km/s$ when combining data for both spacecraft.} Solving Eq.
(\ref{eqn3.12}) for $\delta$, we obtain:%

\begin{align}
j_{P}  &  =\left(  3.85\pm1.88\right)  \times10^{-17}\text{ }cm\ s^{-3}\text{
(Pioneer 10 data)}\label{eqn3.13}\\
\delta &  =\left(  8.12\pm2.35\right)  \times10^{-5}\nonumber\\
j_{P}  &  =\left(  1.37\pm0.95\right)  \times10^{-17}\text{ }cm\ s^{-3}\text{
(error range data)}\nonumber\\
\delta &  =\left(  1.36\pm0.52\right)  \times10^{-4}\nonumber
\end{align}
and these values for $\delta$ are very close to the one we obtained in Ref.
\cite{Varieschi:2008va} ($\delta_{0}=3.83\times10^{-5}$), which was based
solely on the analysis of type-Ia Supernovae data.

\begin{figure}[ptb]%
\centering
\fbox{\ifcase\msipdfoutput
\includegraphics[
width=\textwidth
]
{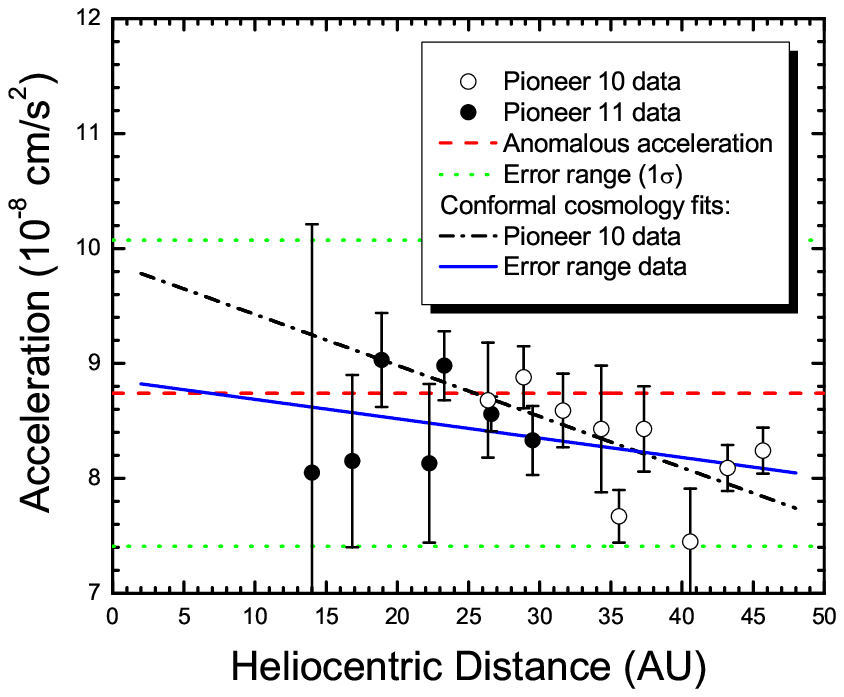}%
\else
\includegraphics[
height=5.152in,
width=6.3023in
]%
{C:/swp55/Docs/KINEMATICAL3/arXiv/graphics/Figure3__3.pdf}%
\fi
}\caption[Early data for Pioneer 10/11 acceleration as a function of
heliocentric distance.]{Early data for Pioneer 10/11 acceleration as a
function of heliocentric distance. The average value of the anomalous
acceleration is indicated in red-dashed, together with its error range
(green-dotted). We also show full conformal cosmology fits of the data, which
allow for a better determination of our cosmological parameters $\gamma$ and
$\delta$.}%
\label{fig3}%
\end{figure}

Another type of analysis is illustrated in Fig. \ref{fig3}. The Pioneer 10/11
data, the standard value of $\left\vert a_{P}\right\vert $ and the related
error range are the same as in the previous figure, but this time we used the
generalized expression of $a_{P}$ in Eq. (\ref{eqn3.11}) to fit the data
within the error range. We allowed both quantities $\gamma$ and $\delta$ to be
free parameters in our fitting procedure and we converted the elapsed time
$t^{\prime}$ in Eq. (\ref{eqn3.11}) into the radial distance $r$ by using the
approximation $r\simeq v_{P}t^{\prime}$, where $v_{P}$ is the average Pioneer
speed, as it was done also for the data in the previous figure. The radial
distance $r$ should be more properly identified with the geocentric distance
of the spacecraft, rather than the heliocentric one, since $r$ should be the
distance from the Earth observer. We also performed fits using the geocentric
distance, but the results were very similar to those obtained by using
heliocentric distances, so we will not include them in the following analysis.%

Again, in Fig. \ref{fig3} we used the expression in Eq. (\ref{eqn3.11}) to fit
the data, although the fitting curves appear almost as straight lines in this
figure. The first conformal cosmology fit, illustrated by the black
(dash-dotted) curve, was obtained by using only the Pioneer 10 data and
yielded the following values of the parameters:%

\begin{align}
\delta &  =\left(  9.19\pm1.53\right)  \times10^{-5}\text{ (Pioneer 10
data)}\label{eqn3.14}\\
\gamma &  =\left(  2.20\pm0.18\right)  \times10^{-28}\text{ }cm^{-1}.\nonumber
\end{align}
The second fit (blue-solid curve) was obtained by using all the data within
the error range (again omitting the first three Pioneer 11 data points) and
produced the following results:%

\begin{align}
\delta &  =\left(  1.38\pm0.43\right)  \times10^{-4}\text{ (error range
data)}\label{eqn3.15}\\
\gamma &  =\left(  1.97\pm0.08\right)  \times10^{-28}\text{ }cm^{-1}.\nonumber
\end{align}

Comparing the results in the last two equations with those for $\delta$\ in
Eq. (\ref{eqn3.13}), obtained with a fixed $\gamma=1.94\times10^{-28}%
\ cm^{-1}$ as in Eq. (\ref{eqn3.10}), we can see that all the values of our
parameters are in agreement. In particular, from the different analyses we
consistently obtain $\gamma\simeq1.9-2.2\times10^{-28}\ cm^{-1}$ and
$\delta\sim10^{-4}-10^{-5}$, where the different values depend on the Pioneer
data being used. As already remarked, the values for $\delta$ quoted above are
also very close to the one we obtained in Ref. \cite{Varieschi:2008va}
($\delta_{0}=3.83\times10^{-5}$), based on type-Ia Supernovae data. In the
next section we will discuss our results and compare them to the current
limits of standard gravity in the solar system.

\section{\label{sect:discussion_conclusions}Discussion of our results and
conclusions}

In the previous sections we discussed how conformal cosmology provides a
natural explanation for the Pioneer anomalous acceleration, in both magnitude
and direction (i.e., the negative sign of the radial acceleration). We also
explained the \textquotedblleft numerical coincidence,\textquotedblright%
\ connecting $a_{P}$ with the Hubble constant, and the observed decrease with
heliocentric distance of $\left\vert a_{P}\right\vert $, related to the
Pioneer jerk $j_{P}$. Although the Pioneer data are still not very accurate,
our analysis consistently indicated that our conformal parameters are
approximately given by $\gamma\sim10^{-28}\ cm^{-1}$ and $\delta\sim
10^{-4}-10^{-5}$ (see Eqs. (\ref{eqn2.9}), (\ref{eqn3.10}) and (\ref{eqn3.13}%
)-(\ref{eqn3.15})). In this final section we will discuss the implications of
the values of our parameters in relation to other studies in the field.

We first remark that a new analysis of rotational velocity data for spiral
galaxies, based on conformal gravity, has recently appeared
\cite{Mannheim:2010ti}, improving the original work on the subject
(\cite{Mannheim:1992vj}, \cite{Mannheim:1996rv}). This new study uses the full
line element of conformal gravity in Eqs. (\ref{eqn2.1})-(\ref{eqn2.2}),
including the effects of the quadratic term $-\kappa r^{2}$, which were
previously neglected, thus obtaining a global gravitational potential
$V_{global}(r)=\frac{\gamma}{2}c^{2}r-\frac{\kappa}{2}c^{2}r^{2}$ of
cosmological origin. In addition to this, a local gravitational potential
$V_{local}(r)$ is obtained by integrating over the visible galactic mass
distribution a gravitational potential per unit solar mass of the form
$V^{\ast}(r)=-G\frac{M_{\odot}}{r}+\frac{\gamma^{\ast}}{2}c^{2}r$. The two
potentials, global and local, are then combined together to model the
rotational motion of galaxies. The fits to galactic rotation data
\cite{Mannheim:2010ti}, performed without any dark matter contribution, show a
remarkable success of conformal gravity, even at the largest distances from
the galactic centers, where the quadratic term $-\kappa r^{2}$ becomes
important and comparable to the linear term $\gamma r$. Mannheim and
collaborators \cite{Mannheim:2010ti}\ were then able to determine the values
of the global universal parameters as $\gamma_{Mann}=3.06\times10^{-30}%
\ cm^{-1}$ and $\kappa_{Mann}=9.54\times10^{-54}\ cm^{-2}$. The related terms
of the global gravitational potential were associated respectively to the
cosmological background and to cosmological inhomogeneities. The local
parameter $\gamma^{\ast}$ was also evaluated as $\gamma^{\ast}=5.42\times
10^{-41}\ cm^{-1}$.

The values of the dimensionful parameters $\gamma$ and $\kappa$ obtained
through this analysis of galactic rotation curves are somewhat different from
our values, reported in this paper or in our previous work
\cite{Varieschi:2008va} ($\gamma_{0}=1.94\times10^{-28}\ cm^{-1}$ and
$\kappa_{0}=6.42\times10^{-48}\ cm^{-2}$). This difference could be due, as we
explained in Ref. \cite{Varieschi:2008va}, to a possible redefinition of the
luminosity distance and other distance indicators, which might affect even the
radial distances (from the galactic centers) which are employed in the
galactic rotation analysis.

However, it is instructive to compute the dimensionless $\delta$ parameter,
using Mannheim's values $\gamma_{Mann}$ and $\kappa_{Mann}$, because this
dimensionless constant should not be affected by a revision of the
cosmological distances. As explained at the beginning of Sect.
\ref{sect:conformal_cosmology}, the parameters $k$, $\gamma$, $\kappa$ and
$\delta$\ are related through $k=-\frac{\gamma^{2}}{4}-\kappa$ and also
$\delta=(\gamma/2\sqrt{\left\vert k\right\vert })$ so that we obtain:%

\begin{align}
k_{Mann}  &  \simeq-\kappa_{Mann}=-9.54\times10^{-54}\ cm^{-2}\label{eqn4.1}\\
\delta_{Mann}  &  =4.95\times10^{-4}.\nonumber
\end{align}
Therefore, the conformal gravity analysis by Mannheim and collaborators
suggests a $\mathbf{k}\equiv\frac{k}{\left\vert k\right\vert }=-1$ Universe,
consistent with our cosmological model and also a value of $\delta
_{Mann}=4.95\times10^{-4}$, close to our quoted values of $\delta\sim
10^{-4}-10^{-5}$.

Conformal gravity considers local gravitational effects as being due to the
local potential $V_{local}$, or simply to the potential $V^{\ast}%
(r)=-G\frac{M_{\odot}}{r}+\frac{\gamma^{\ast}}{2}c^{2}r$ for our solar system.
Since the value of the local constant $\gamma^{\ast}\sim10^{-41}\ cm^{-1}$ is
very small, the modifications to standard dynamics of the solar system are
negligible.\footnote{For example, the ratio between the conformal
gravitational potential $\frac{\gamma^{\ast}}{2}c^{2}r$ and the standard
Newtonian term $G\frac{M_{\odot}}{r}$ at a heliocentric distance of $1\ AU$ is
$\sim10^{-20}$, while at a distance of $100\ AU$ (outer solar system) the same
ratio is $\sim10^{-16}$. Therefore, the \textquotedblleft conformal gravity
force\textquotedblright\ is negligible, compared to the standard Newtonian
one, over the whole solar system region.} Therefore, conformal gravity is not
in any way in contradiction with the very stringent limits on alternative
gravity theories imposed by studies of planetary ephemerides, or other solar
system observations (\cite{Lammerzahl:2006ex}, \cite{2008AIPC..977..254S},
\cite{2010IAUS..261..179S}, \cite{Fienga:2009ub}).

As for our analysis of the Pioneer anomaly, we used the reported values of the
anomalous acceleration $a_{P}$ to determine the cosmological parameters,
simply because such was the way these data were reported in the literature
cited. However, it should be clear from the discussion in Sect.
\ref{sect:pioneer_anomaly} that we explain the Pioneer anomaly in terms of our
\textit{cosmological-gravitational blueshift}, based on the global values of
the parameters $\gamma$, $\kappa$ and $\delta$. In this view, there is no real
dynamic acceleration of the Pioneer spacecraft (or of any other object in the
solar system) oriented toward the Sun, due to some new gravitational force or
modification of existing gravity, except for the tiny corrections coming from
local conformal gravity mentioned above. In fact, in our analysis we assume
that there is no difference between the two velocities $v_{\operatorname{mod}%
}(t^{\prime})$ and $v_{obs}(t^{\prime})$ in Eqs. (\ref{eqn3.3}) and
(\ref{eqn3.5}), therefore the anomalous acceleration defined as $a_{P}%
\equiv\frac{d\left(  \Delta v\right)  }{dt^{\prime}}\simeq\left[  \Delta
v(t^{\prime}+\Delta t^{\prime})-\Delta v(t^{\prime})\right]  /\Delta
t^{\prime}$ with $\Delta v=v_{obs}-v_{\operatorname{mod}}$ is actually zero.

In this way we also overcome the objection, reported in Ref.
\cite{Anderson:2001sg}, that \textquotedblleft the anomalous acceleration is
too large to have gone undetected in planetary orbits, particularly for Earth
and Mars,\textquotedblright\ since \textquotedblleft NASA's Viking mission
provided radio-ranging measurements \cite{1979ApJ...234L.219R}\ to an accuracy
of about $12\ m$,\textquotedblright\ which should have shown the effect of the
anomalous acceleration on the orbits of these two planets.

In our view, precision ranging measurements with radio signals or lasers,
based on the round-trip travel time from Earth to other bodies in the solar
system, would not show any anomalous effect because the speed of light is not
affected by our cosmological model and the corrections to the dynamics of the
solar system due to conformal gravity are negligible.

On the contrary, we would observe an effect similar to the anomalous
acceleration for a spacecraft, a planet, or any other object in the solar
system, if we were to study its motion through Doppler frequency ranging,
because of the intrinsic differences in frequency or wavelength for light
emitted at different spacetime positions, due to our cosmological model.

The size of the local blueshift region, which in our model is responsible for
the frequency differences, can be easily estimated by using Eq.
(\ref{eqn2.8.1}) and the values of our parameters. For example, using the
values from our conformal cosmology fits in Eqs. (\ref{eqn3.14}) and
(\ref{eqn3.15}), we obtain $r_{rs}\simeq50-126\ pc$, corresponding to a
distance comparable to the one between Earth and the nearest bright stars
(which is about $15-30\ pc$). This blueshift region would extend well beyond
the solar system, but would cover a small portion of our galaxy, since
$r_{MilkyWay}\simeq14.6\ kpc$.

The maximum blueshift effect would be seen at $r=\frac{1}{2}r_{rs}%
\simeq25-63\ pc$ and would correspond to a $z_{\min}=\sqrt{1-\delta^{2}}%
-1\sim-10^{-8}$, a very small value. Therefore, the blueshift region and the
related effects are so small that they cannot be practically observed in the
radiation spectrum of stars or other radiation emitting objects within this
region. These effects are only small corrections to the Doppler signals coming
from the Pioneer or other similar spacecraft.

Finally, we want to compare our estimates of the rate of change of the
anomalous acceleration (i.e., the jerk $j_{P}$) with those presented by
independent verifications of the Pioneer anomaly (see review in Ref.
\cite{Turyshev:2010yf}). The first of these studies was performed by Markwardt
\cite{Markwardt:2002ma}, who reviewed data for Pioneer 10 and reported
$a_{P10}=-(7.70\pm0.02)\times10^{-8}\ cm/s^{2}$, with $j_{P10}<0.18\times
10^{-8}\ cm/s^{2}/year=5.70\times10^{-17}\ cm/s^{3}$.\footnote{We prefer to
report here, as also done in the rest of the paper, the anomalous acceleration
$a_{P}$ as a negative quantity and the related jerk $j_{P}$ as a positive
quantity. Some of the papers in the literature adopt the opposite sign
convention, which might generate some confusion.} Using Markwardt values in
Eqs. (\ref{eqn3.9}) and (\ref{eqn3.12}) we obtain $\gamma_{Mark}%
=1.71\times10^{-28}\ cm^{-1}$ and $\delta_{Mark}=5.89\times10^{-5}$,
consistent with our values in Eq. (\ref{eqn3.10}) and Eqs. (\ref{eqn3.13}%
)-(\ref{eqn3.14}) for Pioneer 10.

The second independent study was done by Toth \cite{Toth:2009hx} and reported
results separately for the two spacecraft. From Toth's results for Pioneer 10
($a_{P10}=-(10.96\pm0.89)\times10^{-8}\ cm/s^{2}$, $j_{P10}=(0.21\pm
0.04)\times10^{-8}\ cm/s^{2}/year=6.65\times10^{-17}\ cm/s^{3}$) we compute
$\gamma_{TothP10}=2.44\times10^{-28}\ cm^{-1}$ and $\delta_{TothP10}%
=7.76\times10^{-5}$. Using instead Toth's results for Pioneer 11
($a_{P11}=-(9.40\pm1.12)\times10^{-8}\ cm/s^{2}$, $j_{P11}=(0.34\pm
0.12)\times10^{-8}\ cm/s^{2}/year=1.08\times10^{-16}\ cm/s^{3}$) we obtain
$\gamma_{TothP11}=2.09\times10^{-28}\ cm^{-1}$ and $\delta_{TothP11}%
=5.23\times10^{-5}$, and all these results are also consistent with those
discussed in Sect. \ref{sect:pioneer_anomaly}.

In conclusion, the detailed analysis of the Pioneer anomaly presented in this
work has indicated that our conformal cosmology might be the origin of this
effect, while conformal gravity alone cannot account for the anomalous
acceleration of the spacecraft. If our analysis is correct, it explains
naturally the numerical coincidence between the Pioneer acceleration and the
Hubble constant, including the signs of these quantities. In addition, we
confirm our previous evaluations of the cosmological parameters, $\gamma
_{0}=\left(  1.94\pm0.30\right)  \times10^{-28}\ cm^{-1}$ and $\delta
_{0}=3.83\times10^{-5}$, also in agreement with independent evaluations.
Further studies will be needed when the re-analysis of all the historical
navigational data for the Pioneer spacecraft will be completed by S. Turyshev
and collaborators and new data will be publicly available.

\begin{acknowledgments}
This work was supported by a grant from the Frank R. Seaver College of Science
and Engineering, Loyola Marymount University. The author would like to thank
Dr. S. Turyshev for very useful discussions and advice on the subject.
\end{acknowledgments}

\bibliographystyle{apsrev}
\bibliography{CMB,CONFORMAL,CONSTANTS,COSMOBOOKS,HSTKEY,MOND,OTHERCMB,PIONEER,PK,SDSS,SUPERNOVAE}

\end{document}